\documentclass[10pt,aps,prd,twocolumn,floats,floatfix,showpacs,superscriptaddress,nofootinbib]{revtex4-1}

\usepackage[usenames]{xcolor}
\usepackage{amsmath,amssymb}
\usepackage{mathrsfs}
\usepackage{amsthm}
\usepackage{tikz}
\usetikzlibrary{calc}
\usetikzlibrary{fadings,arrows,shapes,positioning,decorations,intersections}
\usetikzlibrary{patterns}

\definecolor{Green}{rgb}{0.10,0.60,0.30}

\newcommand{\scrI}{\mathscr{I}}

\begin{document}

\title{Dynamical obstruction to perpetual motion from Lorentz-violating black holes}

\author{Robert Benkel}
\email{robert.benkel@nottingham.ac.uk}
\affiliation{School of Mathematical Sciences, University of Nottingham,
University Park, Nottingham, NG7 2RD, UK}

\author{Jishnu Bhattacharyya}
\email{jishnub@gmail.com}
\affiliation{School of Mathematical Sciences, University of Nottingham,
University Park, Nottingham, NG7 2RD, UK}

\author{Jorma Louko}
\email{jorma.louko@nottingham.ac.uk}
\affiliation{School of Mathematical Sciences, University of Nottingham,
University Park, Nottingham, NG7 2RD, UK}

\author{David Mattingly}
\email{david.mattingly@unh.edu}
\affiliation{Department of Physics, University of New Hampshire, Durham, NH 03824, USA}

\author{Thomas P.~Sotiriou}
\email{thomas.sotiriou@nottingham.ac.uk}
\affiliation{School of Mathematical Sciences, University of Nottingham,
University Park, Nottingham, NG7 2RD, UK}
\affiliation{School of Physics and Astronomy, University of Nottingham,
University Park, Nottingham, NG7 2RD, UK}

\date{March 2018; revised June 2018}

\begin{abstract}
Black holes in Lorentz-violating theories have been claimed to violate
the second law of thermodynamics by perpetual motion energy
extraction. We revisit this question for a Penrose splitting process
in a spherically symmetric setting with two species of particles that
move on radial geodesics that extend to infinity. We show that energy
extraction by this process cannot happen in any theory in which
gravity is attractive, in the sense of a geometric inequality that we
describe. This inequality is satisfied by all known Einstein-\ae{}ther
and Ho\v{r}ava black hole solutions.
\end{abstract}
\maketitle

\section{Introduction}

The defining property of a black hole is that nothing can escape from its interior which, in general relativity, is separated from the exterior
by a null hypersurface called the event horizon.
Although the event horizon is a global concept, its nullness implies that local Lorentz symmetry plays a crucial role in the definition of black holes. For if some field excitations propagate superluminally,
then these excitations are not confined by light cones and can
penetrate null event horizons.

There are at least two distinct ways to introduce superluminality.
First, one could have higher-order dispersion relations, such as $\omega^2\propto k^2+a^2 k^4 +\ldots$,
where $\omega$ is the frequency and $k$ the wavenumber, and an unbounded maximum propagation speed.
This scenario occurs in preferred foliation theories
(c.f. \cite{Horava:2009uw,Blas:2009qj,Sotiriou:2009gy,Sotiriou:2009bx,Sotiriou:2010wn}).
Remarkably, the concept of a black hole survives due to a new type of horizon, called the {\em universal horizon} \cite{Barausse:2011pu,Blas:2011ni}. See Ref.\ \cite{Bhattacharyya:2015gwa} for a detailed discussion.
Second, each excitation could
keep a linear dispersion relation but with differing propagation speeds.
In this case different massless excitations propagate along the null cones of distinct effective metrics.
This second scenario should still capture some aspects of the low-momentum behaviour present in the first scenario.

In the second scenario,
a stationary black hole can have multiple horizons
that cloak the interior; each is the Killing horizon of a different effective metric,
with the innermost one corresponding to the fastest excitation
\cite{Eling:2006ec,Berglund:2012bu,Barausse:2011pu,Barausse:2012ny,Barausse:2012qh,Sotiriou:2014gna,Bhattacharyya:2014kta,Barausse:2015frm}.
It has been argued that if such stacks of multiple horizons exist,
they can allow a perpetuum
mobile of the second kind, extracting unlimited energy from the black hole
with no change in the hole's entropy
\cite{Dubovsky:2006vk,Eling:2007qd,Jacobson:2008yc}.
This conclusion is troubling as a point of principle, as it
violates the generalised second law of thermodynamics,
which has been a cornerstone of our understanding
of the quantum properties of black holes~\cite{Hawking:1974sw}
(for modern reviews see~\cite{Wald:1999vt,Page:2004xp}).
The conclusion would also be experimentally intriguing.
While observations have provided strong
constraints on Lorentz violations in both matter and gravity,
no observations, including the recent gravitational wave observations~\cite{TheLIGOScientific:2017qsa,Monitor:2017mdv},
significantly constrain the
speed of superluminal gravity polarisations that
correspond to new degrees of freedom (d.o.f.) inherent
in most models~\cite{Sotiriou:2017obf,Gumrukcuoglu:2017ijh}.
There is an underlying reason -
in the limit from a Lorentz-violating theory to general relativity,
the new d.o.f. do not necessarily decouple,
the limit need not be smooth (see {\em e.g.}~\cite{Charmousis:2009tc,Blas:2009yd,Papazoglou:2009fj}),
and the speeds of the new d.o.f. need not approach the speed of light.

The perpetual motion construction in \cite{Dubovsky:2006vk}
employs Hawking radiation,
arguing that the temperature difference between two horizons
allows heat transfer from a cold to a hot reservoir.
The constructions in \cite{Eling:2007qd,Jacobson:2008yc} are classical,
arguing in terms of the energy of a particle that escapes to
infinity from a splitting event between two horizons.
All require that the region
between the two horizons acts like an ergoregion for the slowest excitations.
Moreover, moving from perpetual motion to a violation
of the generalised second law of thermodynamics
assumes that the process does not
increase the black hole's entropy and has no natural endpoint.
The classical constructions of [17, 18] also rely crucially on an external agent who sets the initial
conditions for the process, arguing that the agent's net contribution to the energy budget can be made negligible.
Finally, the arguments are purely kinematical, with the presumed local geometric structure of the ergoregion
unconstrained by the requirement that it must exist as a solution to any otherwise well-behaved dynamical theory.

Here we re-analyze the perpetual motion question for two primary reasons.
First, subtleties about external agents' contributions to energy budgets
have generated unresolved debates of thirty years and counting~\cite{Marolf:2002ay,Brown:2012un}.
These subtleties are a strong reason to bypass the debates and ask what is possible without external agents.
Second, not all Lorentz violating spacetimes may be solutions of a viable gravitational theory, and we should not consider those spacetimes as representative
of Lorentz violating gravity any more than we consider the negative mass Schwarzschild solution to represent general relativity.

Concretely, we consider a Penrose ``one-to-two''
splitting process involving two different particle species.
The two particles move along radial geodesics of two different (effective) metrics. We assume both metrics are static, spherically symmetric
and asymptotically flat, and that the geodesics extend to
infinity.
The energy budget is therefore
unambiguously defined by the Killing energy at  infinity.
We show that energy extraction by this process cannot happen
in any spacetime in which gravity is attractive,
in the sense of a geometric inequality that we state.
This inequality is equivalently a restriction on the configuration
of the Lorentz-symmetry violating field that
defines the preferred frame, and is satisfied in all
Einstein-\ae{}ther and Ho\v{r}ava gravity solutions
that are analytically or numerically known.

Our results do not directly contest the perpetuum mobile constructions of
\cite{Eling:2007qd,Jacobson:2008yc}.
Nonetheless, they do indicate that removing the external agents
from the construction and considering dynamics introduces a significant new obstruction.
We view this result as a strong and qualitatively new argument
against violation of the generalised second law
in Lorentz-violating gravitational theories.

The structure of the paper is as follows:
Section \ref{sec:scenario} recalls the Lorentz-violating
geometric setting of the perpetual motion scenarios
in~\cite{Eling:2007qd,Jacobson:2008yc}, and Section
\ref{sec:extract-vs-thermo} contrasts the thermodynamic
paradoxes in these scenarios with the thermodynamically
non-paradoxical Penrose processes in Einstein gravity.
Our main result, the perpetual motion exclusion criterion,
is presented in Section~\ref{sec:exclusion}.
Section \ref{sec:admission} presents a partial converse,
which demonstrates that the exclusion criterion
is kinematically nontrivial.
Section \ref{sec:discussion} gives a brief summary and discussion.
Technical material is deferred to three appendices.

\section{Scenario for perpetual motion\label{sec:scenario}}

We first recall
the geometric setting of
the perpetual motion scenarios
in~\cite{Eling:2007qd,Jacobson:2008yc}.
We consider Lorentz-noninvariant theories of gravity that contain the metric
$g_{ab}$ and the \ae{}ther~\cite{Jacobson:2000xp} $u^{a}$, a dynamical unit timelike vector field that defines at each point a preferred timelike direction.
Solutions therefore break local Lorentz symmetry.

Given $g_{ab}$ and~$u^{a}$,
we construct new metrics
\begin{align}
\label{metrics}
g^{(i)}_{ab}
:=
- u_{a} u_{b} + c_i^{-2} \left( u_{a} u_{b} + g_{ab} \right)
\,,
\end{align}
where we have adopted the mostly plus sign convention,
$u_{a} := g_{ab} u^{b}$, and $c_i$ are positive constants.
From the unit timelike condition on $u^a$ it follows that
$g^{(i)}_{ab} u^{a} u^{b} = - 1$ and $u_{a} = g^{(i)}_{ab} u^{b}$ for all~$i$.
Examination of the light cones of $g^{(i)}_{ab}$
in the local Lorentz frame in which $u^a = (1,0,0,0)$ shows that $g^{(i)}_{ab}$
has Lorentzian signature and $c_i$ is its speed of light.
Fields obeying hyperbolic field equations in $g^{(i)}_{ab}$ provide
a covariant description of excitations
that propagate at different speeds in preferred frame theories~\cite{Eling:2006ec}.
In the point particle limit, the fields are replaced by causal geodesics in the respective metrics,
and local interactions between the fields
are described as collisions that preserve
the four-momentum
co-vector~\cite{Eling:2007qd},
as we outline in Appendix~\ref{sec:momentum-cons}.

We consider two particle species,
a slower one denoted by~$A$, propagating along geodesics of $g^A_{ab} := g_{ab}$,
and a faster one, denoted by~$B$,  propagating along geodesics of $g^B_{ab}$
given by \eqref{metrics} with $c_B := c>1$.
The absolute maximum speed of propagation is then~$c$.

We assume that $g^A_{ab}$
is a static, spherically symmetric and asymptotically
flat black hole spacetime, such that the staticity Killing vector
$\chi^a$ asymptotes to the standard Minkowski time translation Killing vector at infinity,
satisfying $g^A_{ab} \chi^a \chi^b \to -1$ there,
and that $g^A_{ab} \chi^a \chi^b$ changes sign at the
Killing horizon. We call the future branch of this Killing horizon
the $A$-horizon.
We further assume that $u^a$ is spherically symmetric and commutes with~$\chi^a$,
and that it asymptotes to $\chi^a$ at the infinity.
Then $g^B_{ab}$ is spherically symmetric
and asymptotically flat, $\chi^a$ is a hypersurface-orthogonal Killing vector of~$g^B_{ab}$,
and $\chi^a$ asymptotes to the standard Minkowski time translation
Killing vector of $g^B_{ab}$ at infinity,
satisfying $g^B_{ab} \chi^a \chi^b \to -1$ there.
Note that while both $g^A_{ab}$ and $g^B_{ab}$
are static in at least some neighbourhood of infinity,
the pairs $(g^A_{ab}, u^a)$ and $(g^B_{ab}, u^a)$ are only stationary
since $u^a$ need not be parallel to $\chi^a$.

\begin{figure}[t!]
\centering
\includegraphics[width=0.9\linewidth]{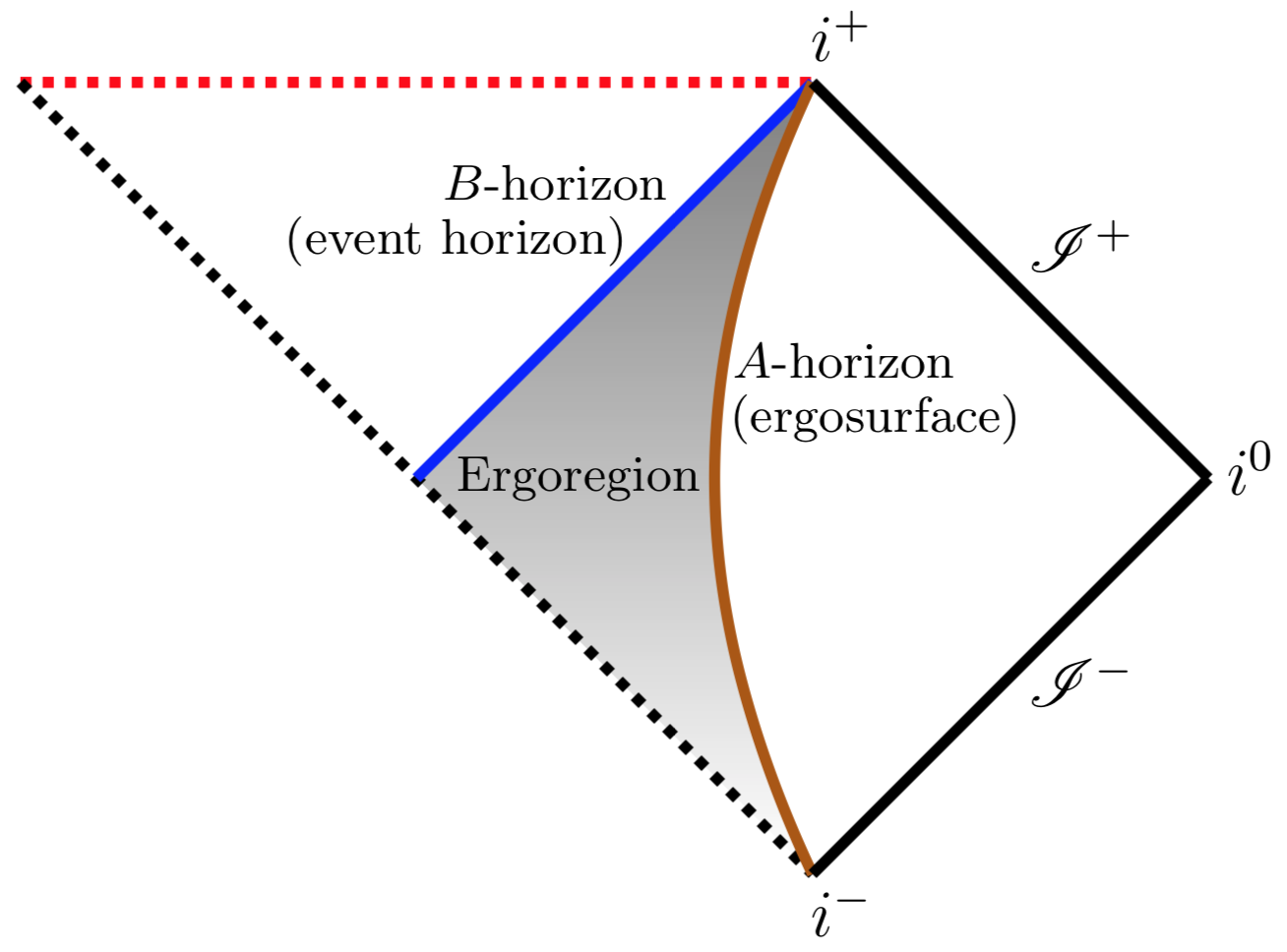}
\caption{Spacetime diagram of the static,
spherically symmetric and asymptotically flat spacetime corresponding
to the effective metric $g^B_{ab}$ of the fastest particle,
in the case where $g^B_{ab}$ describes a black hole
with a single Killing horizon and a spacelike singularity.
The solid brown line marks the
horizon for the $A$ particles,
which can be thought of as an ergosurface.
The solid blue line marks the
horizon for the $B$ particles,
which can be thought of as the event horizon.
The shaded part in between is the ergoregion.
The dotted red line marks the singularity.
The spacetime outside the
$A$-horizon will be referred to
as the outside region and the spacetime inside the
$B$-horizon
as the black hole.\label{fig:spacetime_diagram}}
\end{figure}

We assume that $g^A_{ab}$ and $u^a$ are sufficiently regular
(say, smooth), and in particular that $u^a$ is sufficiently regular
across the $A$-horizon.
As $c>1$, $g^B_{ab} \chi^a \chi^b$ is negative outside and on the $A$-horizon.
Immediately behind
the $A$-horizon there is hence the \emph{ergoregion\/}: a region in which $g^A_{ab} \chi^a \chi^b>0$
but $g^B_{ab} \chi^a \chi^b<0$.
While $g^A_{ab}$ is static only outside the $A$-horizon,
$g^B_{ab}$ is static in the union of the ergoregion, the outside of the $A$-horizon,
and their joint boundary at the $A$-horizon,
with $\chi^a$ providing the timelike hypersurface-orthogonal Killing vector.
The causal structures of $g^A_{ab}$ and $g^B_{ab}$ hence differ:
there exist no causal $A$-curves from the ergoregion to the $\scrI^{+}$ of $g^A_{ab}$,
but there exist causal $B$-curves, and even null $B$-geodesics, from the ergoregion
to the $\scrI^{+}$ of $g^B_{ab}$.
Figure \ref{fig:spacetime_diagram} shows the conformal diagram of $g^B_{ab}$
in the case where
the ergoregion terminates at a $B$-horizon, where $g^B_{ab} \chi^a \chi^b=0=u_a \chi^a$,
and is followed by a region where $g^B_{ab} \chi^a \chi^b>0$
and finally by a spacelike singularity.
It is  worth mentioning that the setup for
$(g^A_{ab}, u^a)$ and $(g^B_{ab}, u^a)$ described above,
though not tied to some specific dynamics, matches exactly that of
known black hole solutions in Einstein-\ae{}ther theory
and Ho\v{r}ava gravity {\em e.g.}~\cite{Eling:2006ec,Barausse:2011pu,Berglund:2012bu}.

Now, as $\chi^{a}$ is $A$-spacelike within the ergoregion,
$A$-particles can carry negative Killing energy there.
This raises the possibility that a system of particles could enter the ergoregion,
undergo  interactions that
create an $A$-particle with negative Killing energy and a $B$-particle
that exits the ergoregion, and, by conservation of total Killing energy,
give the exiting $B$-particle more
Killing energy than originally entered the ergoregion.
This process would thus extract energy from the black hole.

Versions of this process were considered in~\cite{Eling:2007qd,Jacobson:2008yc}.
The process of \cite{Eling:2007qd} introduces an external
agent who releases the initial system
from close to the $A$-horizon, with an initial velocity that is sufficiently
right-pointing in Figure~\ref{fig:spacetime_diagram}.
The process of \cite{Jacobson:2008yc}
introduces an external agent who uses a tether to lower
the initial system from infinity past the $A$-horizon.
In each case there is energy extraction if one can conclusively argue that the agents
and their equipment make a negligible net
contribution to the energy balance.
Here we ask whether the processes have counterparts that
involve no external agents.

\section{Energy extraction versus thermodynamics\label{sec:extract-vs-thermo}}

Before proceeding, we pause to
emphasise that energy extraction
is not on its own a threat to thermodynamics.
For example, for Kerr black holes in Einstein gravity,
energy can be extracted by superradiance
whenever the hole has nonzero angular momentum~\cite{Brito-book};
energy can also be be extracted by the usual Penrose processes
at least for sufficiently rapidly rotating Kerr black holes
\cite{Brito-book,Piran:1977dm,Valles1991,Williams:1995pj,refId0}.
The key point is that the extracted energy does not come for free:
the black hole loses mass by spinning down,
while its horizon and entropy increase.
Moreover, the process has a natural endpoint when black hole has
lost all of its angular momentum~\cite{MTW}. There are no thermodynamical paradoxes.

Could the situation be similar in our case? Suppose for the moment that $g^A_{ab}$ and $g^B_{ab}$ were not related to each other through~\eqref{metrics}, but $g^B_{ab}$ is just some composite metric defined in terms of $g^A_{ab}$ and some other field~$\Psi$. It is a priori conceivable that energy extraction then takes place as in~\cite{Eling:2007qd,Jacobson:2008yc}, and as a result
$\Psi$ reconfigures itself such that eventually $g^B_{ab}$ tends to $g^A_{ab}$, the two horizons merge, the ergoregion disappears and the process halts. Moreover, the entropy never decreases.

However, when $g^A_{ab}$ and $g^B_{ab}$ are related by~\eqref{metrics}, the unit timelike property of $u^a$ shows that
the ergoregion cannot disappear in a regular manner: driving the $B$-horizon towards the $A$-horizon makes the \ae{}ther configuration singular. More broadly, whenever two different excitations propagate at different speeds, which is the hallmark of local Lorentz violation, it is quite difficult to construct a theory where the corresponding horizons merge smoothly.

\section{Perpetual motion exclusion criterion\label{sec:exclusion}}

We now come to the main question of the paper:
when both the initial and final energies are
defined at asymptotically flat infinity,
without external agents,
does energy extraction still occur by some counterpart
of the processes described in~\cite{Eling:2007qd,Jacobson:2008yc}?

We consider a pointlike object $\Sigma$ that
is dropped into the ergoregion from the infinity.
$\Sigma$ can be either an $A$-particle, following a causal $A$-geodesic,
or a $B$-particle, following a causal $B$-geodesic.
In the ergoregion $\Sigma$ splits into
two ejecta, an $A$-particle and a $B$-particle,
each moving on a causal geodesic in their respective metric.
The $B$-ejectum exits the ergoregion and escapes to  infinity.
All the geodesics are assumed to be radial,
in the sense of having vanishing angular momenta with respect
to the Killing vectors of spherical symmetry.
All the particles are assumed to have positive energy in a
local rest frame in the metric whose geodesic they follow.

The back-reaction of the particles on the metric and
the \ae{}ther is neglected throughout. The \ae{}ther, since it has a tensor vacuum expectation value, generates modified light cones for coupled standard model particles in its ground state, in the same way the background tensors in the standard model extension do~\cite{Colladay:1998fq}.
Therefore, the components of the process need not be \ae{}ther excitations
(which take the form of \ae{}ther-metric waves \cite{Jacobson:2004ts}
at the linearized level),
though such excitation could also be used.
For Einstein gravity, the corresponding approximation would mean neglecting the production of gravitational waves.  We stress that this approximation is typical when deriving Lorentz violation constraints from astrophysical observations~\cite{Maccione:2007yc,Liberati:2012jf,Foster:2016uui,Wei:2017zuu}.

Consider now
the asymptotic observers who are static at infinity,
following orbits of the Killing vector~$\chi^a$.
For these observers, the energy of a particle with
the momentum covector $k_a$ is the Killing energy~$- k_a \chi^a$.
The observers see the above process as energy extraction
if and only if the $B$-ejectum has a larger Killing
energy than~$\Sigma$.
Under what conditions on the metrics and
$u^a$ can the asymptotic observers then see energy extraction?

Our main result is the following criterion:

{\bf Perpetual motion exclusion criterion:}
{\it
Suppose that the inequality
\begin{align}
- g^B_{ab}\chi^a \chi^b <1
\label{eq:B-attractivity}
\end{align}
holds everywhere.
Then energy extraction by the process described above cannot happen.}

We sketch the proof here and detail it in Appendix~\ref{sec:thm1proof}.
By spherical symmetry of the geometry and the particle motion,
we may drop the angles and work in the $1+1$ spacetime
dimensions shown in Figure~\ref{fig:spacetime_diagram}.
We raise and lower indices with~$g^B_{ab}$,
and all dot products and normalisations are with respect to~$g^B_{ab}$.

First, we introduce the \ae{}ther frame $(u^{a},s^{a})$,
where the spacelike unit vector $s^{a}$ is orthogonal to $u^{a}$ and tends
to the usual outward-pointing
radial vector at  asymptotically flat infinity.
This frame  makes certain geometric
properties manifest. In particular,
a~covector $k_a$ can be decomposed in the \ae{}ther frame as
\begin{align}
\label{eq:preferred-frame-decomposition_m}
    k_{a}
    =
    E u_{a} + k_s s_{a}
    \,,
\end{align}
and similarly for the corresponding vector~$k^a$.
A~(co)vector with $k_s > 0$ is called right-pointing
and a (co)vector with $k_s < 0$ is called left-pointing.
Also, since the Killing vector $\chi^a$ is
$B$-timelike and future-pointing,
we can parametrise it as
\begin{align}
\label{eq:chi-par-eta_m}
\chi^a = -(u\cdot\chi)(u^a + \tanh\eta\, s^a)
\ ,
\end{align}
where $\eta$ is the rapidity by which $\chi^a$
is boosted relative to~$u^a$.

\begin{figure}[t!]
\centering
\includegraphics[width=0.65\linewidth]{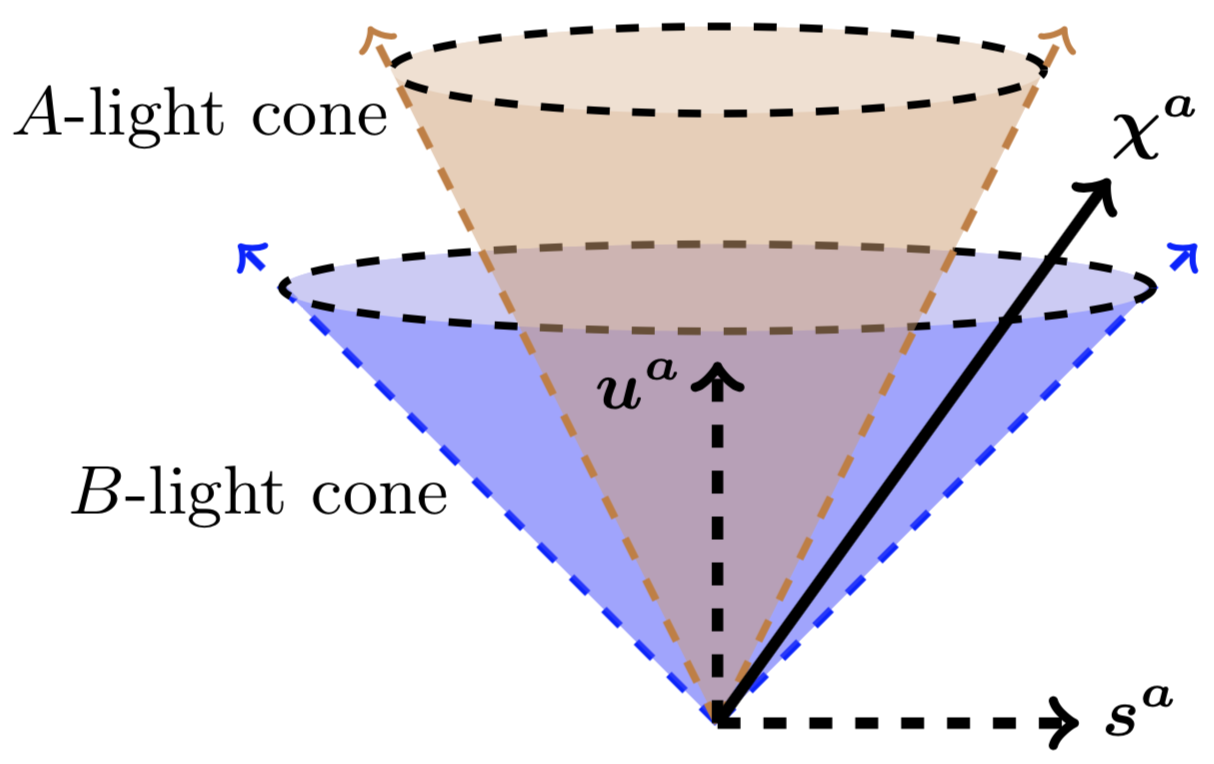}
\caption{The configuration of the two light cones and the Killing vector $\chi^a$
in the ergoregion, in the frame $(u^a,s^a)$.}
\label{pict:ergo_m}
\end{figure}

Now, within the ergoregion, the relative configuration
of $u^a$, $\chi^a$
and the two light cones is as shown in Figure~\ref{pict:ergo_m}.
$\chi^a$~is right-pointing: this follows by continuity because the
$A$-horizon is a future horizon and outside the $A$-horizon
$\chi^a$ is inside the future $A$-light cone.
It follows that
\begin{align}
0 &< s \cdot \chi < - u \cdot \chi
\ .
\label{frame_ineq}
\end{align}

At the splitting event $\Sigma \to A$-ejectum $+$ $B$-ejectum in the ergoregion,
conservation of the four-momentum reads
	\begin{equation}
	\label{eq:4momentum_conservation_m}
	k^{\Sigma}_a = k^{A}_{a} + k^{B}_{a}~,
	\end{equation}
where $k^{\Sigma}_a$, $k^{A}_{a}$ and $k^{B}_{a}$ are the respective momentum
covectors of the three particles.
By conservation of the Killing energy along geodesics in the two metrics,
the energy extraction condition is $-k^{A}_{a}\chi^a<0$.
It follows that $k^{A}_{a}$ is right-pointing, as can be seen by decomposing
$k^{A}_{a}$ as in \eqref{eq:preferred-frame-decomposition_m},
where $E>0$ by the assumption of locally positive energy,
and using~\eqref{frame_ineq}.

It further follows that ${(k^{B})}^{a}$
is more right-pointing than $\chi^a$ in Figure~\ref{pict:ergo_m}.
To see this,  we introduce a function $r$
that is constant on orbits of $\chi^a$
but strictly monotonic from orbit to orbit, increasing towards infinity.
Recall that by assumption the $B$-ejectum makes it to infinity,
where $-\chi\cdot\chi\to1$, and by assumption
$-\chi\cdot\chi<1$ everywhere. The geodesic equation hence implies that the
motion of the $B$-ejectum is strictly monotonic in~$r$,
and by the definition of $r$ this means
${(k^{B})}^{a}$ must be pointing more to the right than~$\chi^a$.
In particular, $k^B_a$ is right-pointing.

We can now turn to~$\Sigma$. By \eqref{eq:4momentum_conservation_m}
and the properties established about $k^{A}_{a}$ and $k^{B}_{a}$,
straightforward geometric considerations show that ${(k^{\Sigma})}^{a}$
must be more right-pointing than ${(k^{B})}^{a}$ in Figure~\ref{pict:ergo_m},
and hence in particular more right-pointing than~$\chi^a$.
Since $\Sigma$ by assumption comes from infinity, and the motion of
a $B$-particle that comes from infinity is strictly monotonic in~$r$,
this shows that $\Sigma$ cannot be a $B$-particle,
and it similarly shows that $\Sigma$ cannot be a massless $A$-particle.
$\Sigma$~must hence be a massive $A$-particle.

Since $\Sigma$ comes from infinity, and Killing energy
is conserved along the trajectory of~$\Sigma$,
we have $- \chi^a k^{\Sigma}_a \ge M$, where $M>0$ is the mass.
Energy extraction requires $-k^{A}_{a}\chi^a<0$.
The remainder of the proof consist of showing that these two inequalities cannot be simultaneously satisfied; the details are in Appendix~\ref{sec:thm1proof}.

Note that any  theory in which gravity as described in the
$B$-metric remains attractive would be expected
to satisfy inequality~\eqref{eq:B-attractivity}. The most obvious cases of theories to consider are Einstein-\ae{}ther theory \cite{Jacobson:2000xp,Jacobson:2008aj} and Ho\v{r}ava gravity \cite{Horava:2009uw,Blas:2009qj,Sotiriou:2010wn} (the latter can be written covariantly in terms of a metric and a hypersurface orthogonal \ae{}ther~\cite{Jacobson:2010mx}).
Under the assumptions of staticity, spherical symmetry, and asymptotic flatness, these two theories have the same black hole solutions \cite{Barausse:2012ny,Barausse:2012qh}. For the two families of analytic solutions given in~\cite{Berglund:2012bu}, inequality \eqref{eq:B-attractivity} does hold: by Eqs.\ (51c), (52), (61) and (63) in~\cite{Berglund:2012bu}, these solutions satisfy
$-u_a \chi^a < 1$, from which \eqref{eq:B-attractivity} follows by~\eqref{eq:chi-par-eta_m}. We have also examined the nine numerical solutions given in~\cite{Barausse:2011pu}, compatible with binary pulsar observational constraints \cite{Foster:2007gr} (although not compatible by the more recent observational
constraints on the speed of gravitational waves \cite{Gumrukcuoglu:2017ijh}),
and a selection of further numerical solutions generated by the code of~\cite{Barausse:2011pu}.
In all cases we find that $-u_a \chi^a < 1$ holds,
implying inequality~\eqref{eq:B-attractivity}.

\section{Perpetual motion admission criterion\label{sec:admission}}

While we have provided evidence that inequality
\eqref{eq:B-attractivity} is a well motivated assumption
in our exclusion criterion,
we now address briefly the converse question: what
would it take to find a configuration
in which energy extraction by our process \emph{does\/} occur?
We show that for any given $A$-metric,
it is kinematically possible to write down \ae{}ther
configurations that allow energy extraction.

If the process involves massive $B$-particles,
relaxing \eqref{eq:B-attractivity}
becomes cumbersome because the qualitative behaviour
of massive $B$-geodesics from and to the infinity is then
sensitive to the locations and heights of the local maxima
of $- g^B_{ab} \chi^a \chi^b$.
However, if the process involves no massive $B$-particles,
we find the following sharp criterion:

{\bf Perpetual motion admission criterion:}
{\it Suppose the process involves no massive $B$-particles.
Then, energy extraction occurs
for some values of the $A$-ejectum mass
if and only if $\Sigma$ is a massive $A$-particle and
the ergoregion has points at which
$K \sqrt{\frac{c+1}{c-1}}< - u_a \chi^a$ and the vector
$V^a := \chi^a - K\sqrt{1 - c^{-2}}\,u^a$ is $B$-timelike,
where $K := \sqrt{\sup \bigl(-g^A_{ab}\chi^a\chi^b\bigr)} \ge1$.}

We give the proof of this criterion in Appendix~\ref{sec:thm2proof}.
We also show there that while
the above statement of the criterion is in terms of
algebraic properties
that are transparently related to the $B$-metric and whose
validity can be readily examined for any given Einstein-\ae{}ther configuration,
these algebraic properties encode a geometric property that
is transparent in terms of the $A$-metric:
energy extraction occurs if and only if, at the splitting event, the \ae{}ther
vector points to the \emph{left\/} of the four-velocity of $\Sigma$
in Figure~\ref{fig:spacetime_diagram},
by an $A$-boost whose velocity is greater than~$c^{-1}$.
The constant $K$ encodes the initial speed with which $\Sigma$
needs to be released from the infinity in order to reach the ergoregion;
in particular, if $-g^A_{ab}\chi^a\chi^b<1$ everywhere, we have $K=1$,
and the optimal case for energy extraction is to release
$\Sigma$ with vanishing initial speed.

This geometric view makes it plain that
for any given $A$-metric, it is possible to write down an \ae{}ther
configuration for which energy extraction occurs:
you just need to make the \ae{}ther sufficiently left-pointing
somewhere in the ergoregion in Figure~\ref{fig:spacetime_diagram}.
These \ae{}ther
configurations will however necessarily
violate~\eqref{eq:B-attractivity},
and we show in Appendix \ref{sec:thm2proof}
that the violation occurs with a wide margin.

Related geometric observations were
used in \cite{Eling:2007qd} to instruct an external
agent how to release an incoming particle from near the $A$-horizon.

\section{Discussion\label{sec:discussion}}

We have analysed classical extraction of energy
from Lorentz-violating black holes in a Penrose splitting process which,
if dynamically possible,
would challenge the status of Lorentz-violating black holes as
thermodynamical objects as implied by the Hawking effect.
The main outcome was that such Penrose processes
do not happen in any black hole configurations in which gravity remains attractive,
in the kinematical sense of inequality~\eqref{eq:B-attractivity},
and we verified that they indeed do not happen in
known analytical or numerical Lorentz-violating black hole solutions.

A~central piece of input was that we considered energy extraction
as seen by observers at an asymptotically flat infinity,
without external agents operating somewhere in the spacetime.
Our notion of energy is hence directly the Killing energy at  infinity,
which should be related to the conserved
charges determined by the dynamics of the theory.
Despite the limitations of spherical symmetry,
the test particle approximation, and the limited number of
particles and
particle species considered,
we view our results as a strong and qualitatively new argument
against violation of the second law of black hole thermodynamics
in theories of gravity without local Lorentz symmetry.
Suggestions that the second
law might be violated
will now face the burden to explain
how our obstruction is avoided.

That being said, suppose that there did exist a
Lorentz-violating theory with a
black hole that allows
energy extraction by our process.
Where would the energy be coming from?
While our test particle approximation
cannot address this question directly,
we saw that energy extraction requires a bump in
$- g^B_{ab}\chi^a \chi^b$. This suggests
that the extraction will affect the bump in some way,
and the ultimate fate of the bump will be related to the
stability properties of the theory.
We leave this question as a subject of future work.

\section*{Acknowledgments}

We thank Ted Jacobson and Eric Poisson
for helpful discussions and correspondence.
JB and TPS acknowledge funding from the European Research Council
under the European Union's Seventh Framework Programme
(FP7/2007-2013), ERC grant agreement n.\ 306425
``Challenging General Relativity.''
JL and TPS acknowledge partial support from Science and Technology Facilities Council
(Theory Consolidated Grants ST/J000388/1 and ST/P000703/1).

\appendix

\section{Momentum co-vector conservation at the splitting\label{sec:momentum-cons}}

In this appendix we justify the momentum co-vector
conservation at the collision event.

Suppose first that $g^A_{ab}$ is Minkowski and $u^a$ is a constant vector.
By \eqref{metrics}, $g^B_{ab}$ is then also Minkowski,
albeit with wider light cones, and
all the translational Killing vectors of $g^A_{ab}$
are also Killing vectors of $g^B_{ab}$. By Noether's theorem,
an action built from fields $\phi^{(j)}$ and
their derivatives using $g^A_{ab}$ and $g^B_{ab}$
has hence four conserved quantities, corresponding to the
four translational Killing vectors.
When the fields are localised into relativistic point particles,
these four conserved quantities reduce on every
hyperplane spacelike in $g^B_{ab}$
(and hence also spacelike in $g^A_{ab}$)
to the sum of the four-momentum co-vectors
of all the particles. Conservation of the total four-momentum
co-vector at particle collisions, of any types of particles,
hence follows from Noether's theorem whenever the
collisions are a localised limit of the underlying
translationally invariant field theory.

When $g^A_{ab}$ and $u^a$ are arbitrary but still smooth,
an action built from fields $\phi^{(j)}$ and
their derivatives using $g^A_{ab}$ and $g^B_{ab}$ need no longer
have conserved quantities
that would correspond to translations.
However, a particle collision process
localised to a single spacetime event
still conserves the four-momentum co-vector,
as is seen by
applying Noether's theorem in a local inertial
frame in a neighbourhood of the collision event.

\section{Proof of the perpetual motion exclusion criterion\label{sec:thm1proof}}

In this appendix we give the detailed proof
of the perpetual motion exclusion criterion
stated in Section \ref{sec:exclusion} of the main text.

\subsection{Notation\label{subsec:thm1proof-notation}}

Recall from Eq.~\eqref{metrics} of the main text
that the $A$-metric and the $B$-metric
are related by
\begin{align}
\label{eq:A-versus-B-metric}
g^B_{ab}
=
- u_{a} u_{b} + c^{-2} \left( u_{a} u_{b} + g^A_{ab} \right)
\ ,
\end{align}
where $u_a$ is unit timelike both in $g^A_{ab}$ and~$g^B_{ab}$, and $c>1$.
We use a
notation adapted to the $B$-metric:
unless otherwise specified, indices are raised and lowered with $g^B_{ab}$,
and dot products and normalisations are with respect to~$g^B_{ab}$.
Given the spherical symmetry of the geometry and of the particle motion,
we may throughout drop the angles and work as if in $1+1$ dimensions.

\subsection{Geometric preliminaries\label{subsec:thm1proof-geometric}}

We need four preliminaries.

First, we introduce the \ae{}ther-adapted frame $(u^{a},s^{a})$,
where the spacelike unit vector $s^{a}$ is
orthogonal to $u^{a}$ and tends to the usual outward-pointing
radial vector at the asymptotically flat infinity. We call this frame the \ae{}ther frame.
A covector $k_a$ can be decomposed in the \ae{}ther frame as
\begin{align}
\label{eq:preferred-frame-decomposition}
    k_{a}
    =
    E u_{a} + k_s s_{a}
    \,,
\end{align}
and similarly for the corresponding vector~$k^a$.
A (co)vector with $k_s > 0$ is called right-pointing
and a (co)vector with $k_s < 0$ is called left-pointing.

Second, since the Killing vector $\chi^a$ is
$B$-timelike and future-pointing,
we can parametrise it as
\begin{align}
\label{eq:chi-par-eta}
\chi^a = -(u\cdot\chi)(u^a + \tanh\eta\, s^a)
\ ,
\end{align}
where $\eta$ is the rapidity by which $\chi^a$
is boosted relative to~$u^a$. The normalisation factor in \eqref{eq:chi-par-eta}
can be verified by contracting both sides with~$u_a$.


Third, within the ergoregion, the relative configuration
of $u^a$, $\chi^a$ and the two light cones
is as shown in Figure~\ref{pict:ergo_m},
in the frame $(u^{a},s^{a})$.
$u^a$~is straight up, within the future $A$-light cone,
while $\chi^a$ is sandwiched between the two future light cones.
$\chi^a$~is right-pointing: this follows by continuity because the
$A$-horizon is a future horizon and outside the $A$-horizon
$\chi^a$ is inside the future $A$-light cone.
It follows that
\begin{align}
0 &< s \cdot \chi < - u \cdot \chi
\ ,
\label{eq:ergoregion_inequality}
\\
1 & < c\tanh\eta
\ .
\label{eq:eta-ineq1}
\end{align}
Introducing now the assumption $-\chi\cdot \chi < 1$,
\eqref{eq:chi-par-eta} implies $-u\cdot\chi < \cosh\eta$,
and combining this with \eqref{eq:ergoregion_inequality} gives
\begin{equation}
0 < -u\cdot\chi < \cosh\eta
\ .
\label{eq:eta-ineq2}
\end{equation}

Fourth, we introduce a function $r$
that is constant on orbits of $\chi^a$
but strictly monotonic from orbit to orbit, increasing towards the infinity.
(For some Einstein-\ae{}ther solutions $r$ may be chosen to be
the area-radius, but we wish to proceed here without
assumptions about the field equations.)


\subsection{Conservation laws\label{subsec:thm1proof-cons}}

Consider now the splitting event $\Sigma \to A$-ejectum $+$ $B$-ejectum
in the process described in the main text.
Conservation of the four-momentum at the splitting event reads
	\begin{equation}
	\label{eq:4momentum_conservation}
	k^{\Sigma}_a = k^{A}_{a} + k^{B}_{a}~,
	\end{equation}
where $k^{\Sigma}_a$, $k^{A}_{a}$ and $k^{B}_{a}$ are the respective momentum
covectors of the three particles.
By conservation of the Killing energy along geodesics in the two metrics,
the energy extraction condition is $-k^{A}_{a}\chi^a<0$.
It follows that $k^{A}_{a}$ is right-pointing, as can be seen by decomposing
$k^{A}_{a}$ as in \eqref{eq:preferred-frame-decomposition},
where $E>0$ by the assumption of locally positive energy,
and using~\eqref{eq:ergoregion_inequality}.

It further follows that ${(k^{B})}^{a}$
is more right-pointing than $\chi^a$ in Figure~\ref{pict:ergo_m}.
To see this, recall that by assumption the $B$-ejectum makes it to the infinity,
where $-\chi\cdot\chi\to1$, and by assumption
$-\chi\cdot\chi<1$ everywhere. The geodesic equation hence implies that the
motion of the $B$-ejectum is strictly monotonic in~$r$,
and by the definition of $r$ this means
${(k^{B})}^{a}$ must be pointing more to the right than~$\chi^a$.
In particular, $k^B_a$ is right-pointing.

We can now turn to~$\Sigma$. By \eqref{eq:4momentum_conservation}
and the properties established about $k^{A}_{a}$ and $k^{B}_{a}$,
straightforward geometric considerations show that ${(k^{\Sigma})}^{a}$
must be more right-pointing than ${(k^{B})}^{a}$ in Figure~\ref{pict:ergo_m},
and hence in particular more right-pointing than~$\chi^a$.
Since $\Sigma$ by assumption comes from the infinity,
and the motion of a $B$-particle that comes from the infinity is strictly monotonic in~$r$,
this shows that $\Sigma$ cannot be a $B$-particle,
and it similarly shows that $\Sigma$ cannot be a massless $A$-particle.
$\Sigma$~must hence be a massive $A$-particle.
We may parametrise $k^{\Sigma}_{a}$ as
\begin{equation}
\label{Sigma=mA}
k^{\Sigma}_a = M (\cosh\beta \, u_a + c\sinh\beta \, s_a)
\ ,
\end{equation}
where $M>0$ is the mass.
Note that ${(g^A)}^{ab} k^{\Sigma}_a k^{\Sigma}_b = - M^2$,
where ${(g^A)}^{ab}$ is the usual inverse of $g^A_{ab}$.
Since $\Sigma$ comes from the infinity,
we have $- \chi^a k^{\Sigma}_a \ge M$, which by \eqref{eq:chi-par-eta} gives
\begin{align}
1 \le -(u\cdot\chi) (\cosh\beta  - c \tanh\eta \sinh\beta)
\ .
\label{eq:Sigma-comes-from-infty}
\end{align}

Finally, consider the ejecta.
We parametrise $k^A_{a}$ and $k^B_{a}$ as
\begin{subequations}
\label{eq:momvecs-para}
\begin{align}
k^A_{a} &= \sqrt{m^2 + p^2}\, u_a + c p s_a
\ ,
\\
k^B_{a} &= \sqrt{\mu ^2 + q^2}\, u_a + q s_a
\ ,
\label{eq:momvecs-para-B}
\end{align}
\end{subequations}
where $m\ge0$ and $\mu\ge0$ are the respective masses. Note that
${(g^A)}^{ab} k^A_a k^A_b = - m^2$ and $k^B \cdot k^B = - \mu^2$.
As ${(k^{B})}^{a}$ is more right-pointing than~$\chi^a$, comparison of
$\eqref{eq:momvecs-para-B}$ and \eqref{eq:chi-par-eta} gives
\begin{align}
\tanh\eta  < \frac{q}{\sqrt{\mu^2 + q^2}}
\ ,
\label{eq:q-rightpointing}
\end{align}
and since the $B$-ejectum reaches infinity, we have
$- \chi^a k^B_a \ge \mu$, or
\begin{align}
\mu \le -(u\cdot\chi) (\sqrt{\mu^2 + q^2}  - q\tanh\eta)
\ .
\label{eq:B-reaches-infty}
\end{align}
The energy extraction inequality, $-k^{A}_{a}\chi^a<0$, reads
\begin{align}
\sqrt{m^2 + p^2}  < c p \tanh\eta
\ .
\label{eq:p-extraction}
\end{align}
Momentum conservation
\eqref{eq:4momentum_conservation} takes the form
\begin{subequations}
\label{eq:momcons-explicit}
\begin{align}
M \cosh\beta &= \sqrt{m^2 + p^2} + \sqrt{\mu ^2 + q^2}
\ ,
\\
M \sinh\beta &= p + q/c
\ ,
\label{eq:momcons-spatial-explicit}
\end{align}
\end{subequations}
and solving this pair for $\beta$ gives
\begin{align}
\tanh\beta = \frac{p + q/c}{\sqrt{m^2 + p^2} + \sqrt{\mu ^2 + q^2}}
\ .
\label{eq:beta-solution}
\end{align}
Note that since $q>0$ by \eqref{eq:q-rightpointing}
and $p>0$ by~\eqref{eq:p-extraction}, \eqref{eq:beta-solution} implies $\beta>0$.

\subsection{Contradiction}

We shall now show that the above set of inequalities has no solutions.


To begin, we note by \eqref{eq:eta-ineq1} that we may
introduce the positive number $\epsilon$ by
\begin{align}
\tanh\epsilon = \frac{1}{c\tanh\eta}
\ .
\label{eq:eps-def}
\end{align}
In terms of~$\epsilon$,
\eqref{eq:p-extraction} becomes
\begin{align}
\tanh\epsilon &< \frac{p}{\sqrt{m^2 + p^2}}
\label{eq:p-extraction-epsilon}
\end{align}
and \eqref{eq:Sigma-comes-from-infty} becomes
\begin{align}
N \le \frac{\sinh(\epsilon-\beta)}{\sinh\epsilon}
\ ,
\label{eq:Sigma-comes-from-infty-epsilon}
\end{align}
where we have written $N := - 1/(u\cdot\chi)$. Since $\beta>0$,
\eqref{eq:Sigma-comes-from-infty-epsilon} shows that $N<1$.
By \eqref{eq:eta-ineq2} we then have
\begin{align}
\frac{1}{\cosh\eta} < N < 1
\ .
\label{eq:eta-N-1-ineq}
\end{align}
Solving \eqref{eq:Sigma-comes-from-infty-epsilon} for $\beta$
gives $0<\beta\le \beta_c$, where
\begin{align}
e^{-\beta_c} = e^{-\epsilon}
\left(
N \sinh\epsilon + \sqrt{{(N\sinh\epsilon)}^2 + 1}
\, \right)
\ .
\label{eq:betac-def}
\end{align}
For fixed~$\epsilon$,
$\beta_c$ is strictly decreasing in~$N$.
From \eqref{eq:eta-N-1-ineq} we then have
$\beta_c < \beta_e$, where
\begin{align}
e^{-\beta_e} = e^{-\epsilon}
\left(
\frac{\sinh\epsilon}{\cosh\eta} + \sqrt{\left(\frac{\sinh\epsilon}{\cosh\eta}\right)^{\!2} + 1}
\, \right)
\ .
\label{eq:betam-def}
\end{align}
Collecting, we have $0<\beta<\beta_e$.


From here on we consider the cases $\mu>0$ and $\mu=0$ in turn.

\subsubsection{$\mu>0$}

Suppose $\mu>0$.
We write $q = \mu \sinh\psi$, where $\psi>\eta$ by~\eqref{eq:q-rightpointing}.

%

If $m>0$, we write $p = m \sinh\theta$.
\eqref{eq:beta-solution} then becomes
\begin{align}
\tanh\beta = \frac{(m/\mu)\sinh\theta + c^{-1} \sinh\psi}{(m/\mu)\cosh\theta + \cosh\psi}
\ .
\label{eq:beta-solution-psi-mpos}
\end{align}
From \eqref{eq:p-extraction} it follows that $\tanh\theta > c^{-1}$,
and using this, an elementary analysis of \eqref{eq:beta-solution-psi-mpos}
shows that
\begin{align}
\tanh\beta > c^{-1}\tanh\psi
\ .
\label{eq:beta-versus-psi}
\end{align}
If instead $m=0$,
\eqref{eq:beta-solution} becomes
\begin{align}
\tanh\beta = \frac{(p/\mu) + c^{-1} \sinh\psi}{(p/\mu) + \cosh\psi}
\ ,
\label{eq:beta-solution-psi-mzero}
\end{align}
leading again to~\eqref{eq:beta-versus-psi}.
We conclude that
\eqref{eq:beta-versus-psi} holds for all $m\ge0$.

Starting from~\eqref{eq:beta-versus-psi}, we have
\begin{align}
c > \frac{\tanh\psi}{\tanh\beta}
> \frac{\tanh\eta}{\tanh\beta_e}
\ ,
\label{eq:c-chain-firstpart}
\end{align}
where the first inequality is \eqref{eq:beta-versus-psi} while the
second inequality uses $\psi>\eta$ and $0<\beta<\beta_e$.
Using \eqref{eq:eps-def} and~\eqref{eq:betam-def}, it can be shown that the rightmost
expression in \eqref{eq:c-chain-firstpart} is equal to
\begin{align}
c + \frac{\sqrt{c^2-1}}{\cosh\eta}
\ ,
\end{align}
which is strictly greater than~$c$, in contradiction
with~\eqref{eq:c-chain-firstpart}.
This completes the proof for $\mu>0$.

\subsubsection{$\mu=0$}

Suppose $\mu=0$.

If $m>0$, we write again $p = m \sinh\theta$.
\eqref{eq:beta-solution} then becomes
\begin{align}
\tanh\beta = \frac{(m/q)\sinh\theta + c^{-1}}{(m/q)\cosh\theta + 1}
\ .
\label{eq:beta-solution-psiinfty-mpos}
\end{align}
As $\tanh\theta > c^{-1}$,
an elementary analysis of \eqref{eq:beta-solution-psiinfty-mpos}
shows that
\begin{align}
\tanh\beta > c^{-1}
\ .
\label{eq:beta-versus-unity}
\end{align}
If instead $m=0$,
\eqref{eq:beta-solution} becomes
\begin{align}
\tanh\beta = \frac{(p/q) + c^{-1}}{(p/q) + 1}
\ ,
\label{eq:beta-solution-psiinfty-mzero}
\end{align}
leading again to~\eqref{eq:beta-versus-unity}.
We conclude that
\eqref{eq:beta-versus-unity}
holds for all $m\ge0$.

If follows that \eqref{eq:c-chain-firstpart} is replaced by
\begin{align}
c > \frac{1}{\tanh\beta}
> \frac{1}{\tanh\beta_e}
\ .
\label{eq:c-masslesschain-firstpart}
\end{align}
As the rightmost expression in \eqref{eq:c-masslesschain-firstpart}
is strictly greater than the rightmost expression in \eqref{eq:c-chain-firstpart},
a contradiction again follows.
This completes the proof for $\mu=0$.
$\blacksquare$

\section{Proof of the perpetual motion admission criterion\label{sec:thm2proof}}

In this appendix we give the proof of the perpetual motion
admission criterion stated in Section \ref{sec:admission} of the main text.
The notation is as in Appendix~\ref{sec:thm1proof}.
As stated in the assumptions of the admission criterion, any
$B$-particles involved in the process are assumed to be massless.

\subsection{Proof}

The geometric preliminaries of Appendix \ref{sec:thm1proof}
hold with one exception:
since the $-\chi\cdot \chi < 1$ assumption has been dropped, the inequality
$-u\cdot\chi < \cosh\eta$ in
\eqref{eq:eta-ineq2} need no longer hold.

Consider the splitting event.
$k^A_a$~and $k^B_a$ are given by~\eqref{eq:momvecs-para},
where now $\mu=0$ by assumption.
The $B$-ejectum reaches infinity iff $q>0$,
and the energy extraction inequality is~\eqref{eq:p-extraction},
implying in particular that $p>0$.

If $\Sigma$ is a massless $B$-particle, $k^\Sigma_a$ is
a positive multiple of $u_a - s_a$, where the coefficient of $s_a$ is negative since $\Sigma$
by assumption comes from the infinity.
Similarly, if $\Sigma$ is a massless $A$-particle, $k^\Sigma_a$ is
a positive multiple of $u_a - c s_a$.
Both of these cases are however inconsistent with the
$s_a$-projection of the momentum conservation equation~\eqref{eq:4momentum_conservation}.
Hence $\Sigma$ is a massive $A$-particle, and $k^\Sigma_a$ is given by \eqref{Sigma=mA} with $M>0$.

The momentum conservation equation \eqref{eq:4momentum_conservation}
becomes \eqref{eq:momcons-explicit} with $\mu=0$.
For $m>0$, momentum conservation is hence equivalent to
\eqref{eq:momcons-spatial-explicit} and~\eqref{eq:beta-solution-psiinfty-mpos},
and the energy extraction condition is
$\tanh\theta > c^{-1}$; similarly, for $m=0$,
momentum conservation is equivalent to
\eqref{eq:momcons-spatial-explicit} and~\eqref{eq:beta-solution-psiinfty-mzero},
and the energy extraction condition is $p>0$.
An elementary analysis shows that a solution with some $m\ge0$ exists iff
\begin{align}
\tanh\beta > c^{-1}
\ .
\label{eq:splitting-relative-velocity-condition}
\end{align}
For given $\beta$ satisfying \eqref{eq:splitting-relative-velocity-condition},
the range of $m$ for which solutions exist always includes $m=0$.

To see the geometric meaning of~\eqref{eq:splitting-relative-velocity-condition},
recall that the four-velocity vector of $\Sigma$
at the splitting event is $v_\Sigma^a := \cosh\beta \, u^a + c^{-1}\sinh\beta \, s^a$,
as can be seen by raising the index of $k^{\Sigma}_b$ with ${(g^A)}^{ab}$ and normalising.
As $(u^a ,  c^{-1} s^a)$ is a normalised frame in  the $A$-metric,
\eqref{eq:splitting-relative-velocity-condition}
says that $v_\Sigma^a$ points to the right of $u^a$
by an $A$-metric boost whose velocity is greater than~$c^{-1}$.

To summarise:
energy extraction occurs for some values of the $A$-ejectum mass
if and only if $\Sigma$ is a massive $A$-particle
and \eqref{eq:splitting-relative-velocity-condition}
holds at the splitting event. Geometrically,
\eqref{eq:splitting-relative-velocity-condition}
says that the four-velocity vector of $\Sigma$
at the splitting event
points to the right of $u^a$
by an $A$-metric boost whose velocity is greater than~$c^{-1}$.

We next need to examine under what conditions $\Sigma$ makes it to the ergoregion.
Recall that $-g^A_{ab}\chi^a\chi^b \to 1$ at the infinity,
and $-g^A_{ab}\chi^a\chi^b$ changes sign at the ergosurface.
It follows that $K := \sqrt{\sup \bigl(-g^A_{ab}\chi^a\chi^b\bigr)}$
exists and satisfies $1\le K < \infty$. The
$A$-metric geodesic equation shows that $\Sigma$ makes it to the ergoregion
iff its initial speed $v\ge0$ at the infinity satisfies
$1/\sqrt{1-v^2} > K$,
except that when $-g^A_{ab} \chi^a \chi^b<1$ everywhere,
$\Sigma$ makes it to the ergoregion also for $v=0$.
The geodesic equation further shows that increasing $v$ makes the four-velocity vector
of $\Sigma$ more left-pointing in the ergoregion.
It follows that the threshold case for energy extraction
is when the momentum covector of the $\Sigma$-trajectory approaches
$M (K u_a - \sqrt{K^2-1} \, c s_a)$ at the infinity.
Killing energy conservation on the threshold case trajectory
gives $- \chi^a k^A_a = MK$, which by
\eqref{eq:chi-par-eta} and \eqref{Sigma=mA} takes the form
\begin{align}
K &= -(u\cdot\chi) (\cosh\beta  - c \tanh\eta \sinh\beta)
\notag
\\
&= -(u\cdot\chi) \frac{\sinh(\epsilon-\beta)}{\sinh\epsilon}
\ ,
\label{eq:thm2:killing}
\end{align}
where in the second equality $\epsilon$ is given by
\eqref{eq:eps-def} and we have used the same rearrangement
as in~\eqref{eq:Sigma-comes-from-infty-epsilon}.
As the last expression in \eqref{eq:thm2:killing} is strictly decreasing in~$\beta$,
the condition \eqref{eq:splitting-relative-velocity-condition}
shows that energy extraction occurs iff
\begin{align}
K < -(u\cdot\chi) \, \frac{(1 - \tanh\eta)}{\sqrt{1 - c^{-2}}}
\ ,
\label{eq:splitting2-relative-velocity-condition}
\end{align}
where we have noted that $\cosh\beta \to 1/\sqrt{1-c^{-2}}$
and $\sinh\beta \to c^{-1} /\sqrt{1-c^{-2}}$ as $\tanh\beta \to c^{-1}$.

To express \eqref{eq:splitting2-relative-velocity-condition}
in terms of invariant quantities, we note from
\eqref{eq:chi-par-eta} that
\begin{align}
- \chi\cdot\chi = {(u\cdot\chi)}^2 \left(1 - \tanh^2\!\eta\right)
\ .
\label{eq:chisquared-vs-eta}
\end{align}
Suppose that \eqref{eq:splitting2-relative-velocity-condition} holds.
As $\tanh\eta > c^{-1}$, \eqref{eq:splitting2-relative-velocity-condition} implies
\begin{align}
K \sqrt{\frac{c+1}{c-1}} < - u\cdot\chi
\ ,
\label{eq:thm2-bottom-direction}
\end{align}
and eliminating $\eta$ with the help of
\eqref{eq:chisquared-vs-eta} and using \eqref{eq:thm2-bottom-direction} shows that the vector
\begin{align}
V^a := \chi^a - K \sqrt{1 - c^{-2}} \, u^a
\label{eq:thm2-B-timelike-vector}
\end{align}
is $B$-timelike. Conversely, if \eqref{eq:thm2-bottom-direction}
holds and $V^a$ is $B$-timelike,
direct algebra using \eqref{eq:chisquared-vs-eta} shows
that \eqref{eq:splitting2-relative-velocity-condition} holds.

This completes the proof.
$\blacksquare$

\subsection{Comments}

We end with two comments.

First, the $B$-timelikeness of $V^a$
says geometrically that the four-velocity vector of $\Sigma$ and the
\ae{}ther are related by an $A$-boost whose speed exceeds~$c^{-1}$,
but the $B$-timelikeness of $V^a$ on its own does not specify the direction of the boost.
The condition \eqref{eq:thm2-bottom-direction} implies $K < - u\cdot\chi$,
which by \eqref{eq:thm2:killing} implies that $\beta>0$, which says that
the four-velocity vector of $\Sigma$ points to the right of the
\ae{}ther. The only role of \eqref{eq:thm2-bottom-direction} is hence to establish the
relative orientation of the two vectors.
\eqref{eq:thm2-bottom-direction} could therefore be replaced by any weaker condition that
does the same, such as $K < - u\cdot\chi$.

Second, an elementary analysis of \eqref{eq:splitting2-relative-velocity-condition}
and \eqref{eq:chisquared-vs-eta} shows that there
exist configurations in which $-\chi \cdot \chi>1$ somewhere in the ergoregion
but \eqref{eq:splitting2-relative-velocity-condition}
nevertheless does not hold, not even with $K=1$.
This shows that the perpetual motion exclusion criterion in the main text cannot be sharp
at least when the processes are assumed to involve no massive $B$-particles.

\bibliography{draft}

\end{document}